\documentclass[aps, pra,superscriptaddress,preprint]{revtex4-2}
\usepackage[cmex10]{amsmath}
\usepackage{amssymb}
\usepackage[english]{babel}
\usepackage{physics}
\usepackage[normalem]{ulem}
\usepackage{graphicx}
\usepackage{float}
\usepackage[hidelinks]{hyperref}
\usepackage{glossaries}
\usepackage{siunitx}
\usepackage{xcolor}
\usepackage{bm}
\usepackage{bbm}
\usepackage{color}
\DeclareMathAlphabet{\mathcal}{OMS}{cmsy}{m}{n}

\addto\captionsenglish{%
  
}

\begin{document}
\title{Topologically protected frequency control of broadband signals\\ in dynamically modulated waveguide arrays}

\author{Francesco S. Piccioli}
\affiliation{INO-CNR BEC Center and Dipartimento di Fisica, Università di Trento, 38123 Trento, Italy}
\affiliation{Institute for Physics, University of Rostock, Albert-Einstein-Str. 23, 18059 Rostock, Germany}
\email{francesco.piccioli@unitn.it}

\author{Alexander Szameit}
\affiliation{Institute for Physics, University of Rostock, Albert-Einstein-Str. 23, 18059 Rostock, Germany}

\author{Iacopo Carusotto}
\affiliation{INO-CNR BEC Center and Dipartimento di Fisica, Università di Trento, 38123 Trento, Italy}

\begin{abstract}
We theoretically propose a synthetic frequency dimension scheme to control the spectrum of a light beam propagating through an array of evanescently coupled waveguides modulated in time by a propagating sound wave via the acousto-optical effect. Configurations are identified where the emerging two-dimensional synthetic space-frequency lattice displays a non-trivial topological band structure. The corresponding chiral edge states can be exploited to manipulate the frequency spectrum of an incident beam in a robust way. In contrast to previous works, our proposal is not based on discrete high-Q cavity modes and can be applied to the manipulation of broadband signals with arbitrary spectra.
\end{abstract}

\maketitle
\section{introduction}
The control of the frequency spectrum of a light beam is of high importance in a number of photonic applications, from high speed communications to quantum technologies. In telecommunications, the concept of Frequency Division Multiplexing (FDM) is at the basis of 5G technology and allows to increase the data throughput by superposing in time signals with different carrier frequencies \cite{Swaminathan2017}. In quantum optics, entanglement and quantum interference processes involving the frequency and/or temporal degrees of freedom are attracting a strong interest~\cite{pe2005temporal,kobayashi2016frequency,roslund2014wavelength} and strategies for a careful compensation of frequency mismatch between different emitters are often required~\cite{Vural2020, Colautti2020, Liu2020}.

In all these applications, the ability of manipulating the spectrum of generic signals 
is a fundamental basic block in the direction of building complex networks.
Among the many processes that can be exploited to this purpose, such as nonlinear optical wave-mixing~\cite{Xu1993,Hiroki2008, Bell2017,Weber2019, Li2019} or photon energy lifter~\cite{Gaburro2006,Preble2007}, a most intriguing one is undoubtedly the dynamical modulation in time of the optical properties of a medium. Such mechanism has been recently exploited to realize synthetic dimensions~\cite{Ozawa2016,Yuan2016,Yuan2018_Optica} in which the frequency degree of freedom plays the role of an extra physical dimension. This leads to an effective description of light propagation in terms of a higher-dimensional wave equation, where different kind of forces can be engineered to control the dynamics in the full high-dimensional space.

Most of the so far proposed synthetic frequency dimension schemes rely on the dynamical modulation of resonating structures like multi-mode cavities~\cite{Yuan2021,Buddhiraju21}, where high quality factors are typically needed. Not only does this impose stringent limits on the fabrication quality, but also restricts the signals to be manipulated to quasi-monochromatic light beams or frequency combs with a spacing commensurate to the free spectral range of the resonators. The severity of this restriction is most apparent when contrasted to the typically broadband nature of telecommunication signals and to the inhomogeneous broadening of quantum emitters. 

This limitation can be overcome by replacing resonating configurations with propagating ones such as optical waveguides. Recent works have started exploring the temporal modulation of isolated LiNbO{\scriptsize 3} optical waveguides~\cite{Qin2018_PRA,Qin2018_PRL,Qin2020} using an RF-driven electro-optical effect. However, many interesting phenomena such as topological effects \cite{Ozawa2016, Qin2018_OptExpr, Ozawa2019, Leykam2020} requires at least another dimension. Beside using multi-frequency temporal modulations~\cite{Yuan2018_PRB}, reciprocal space schemes~\cite{Lustig2019,Nemirovsky2021,Ozawa2021}, or spin-like degrees of freedom~\cite{Dutt2020}, a most direct way to increase the dimensionality is to use evanescently Coupled Waveguide Arrays (CWA)~\cite{Szameit2010}. 
A pioneering proposal in this direction was recently put forward, based on a complicated lattice geometry~\cite{wu2021quasi}.

In this article, we propose the use of propagating sound waves to dynamically modulate a simpler uniformly-spaced CWA via the acousto-optic effect. This allows to realize a hybrid synthetic space composed by a spatial and a frequency dimension and, thus, observe topological photonics effects of great interest for applications. While our proposal directly extends to electro-optically modulated CWAs, the use of sound-induced acousto-optical modulation in laser-written amorphous materials rather than the electro-optic effect in patterned crystalline materials eases fabrication. Moreover, the larger inter-waveguide distances of laser-written devices is compatible with the long wavelength of the sonic waves considered in this work. As a result, one can make use of standard photonic technology to design a suite of efficient schemes that take advantage of topological protection for the robust coherent manipulation of the carrier and/or the linewidth of broadband signals. A crucial further theoretical step beyond the recent work~\cite{wu2021quasi}, we explicitly show how our proposal is able to efficiently manipulate broadband beams with arbitrary spectra other than a comb of monochromatic components, including single optical pulses. Furthermore, we highlight how our proposal does not require additional waveguides to selectively inject light into the desired mode of the lattice.
\begin{figure}
    \centering
    \includegraphics[scale = 1]{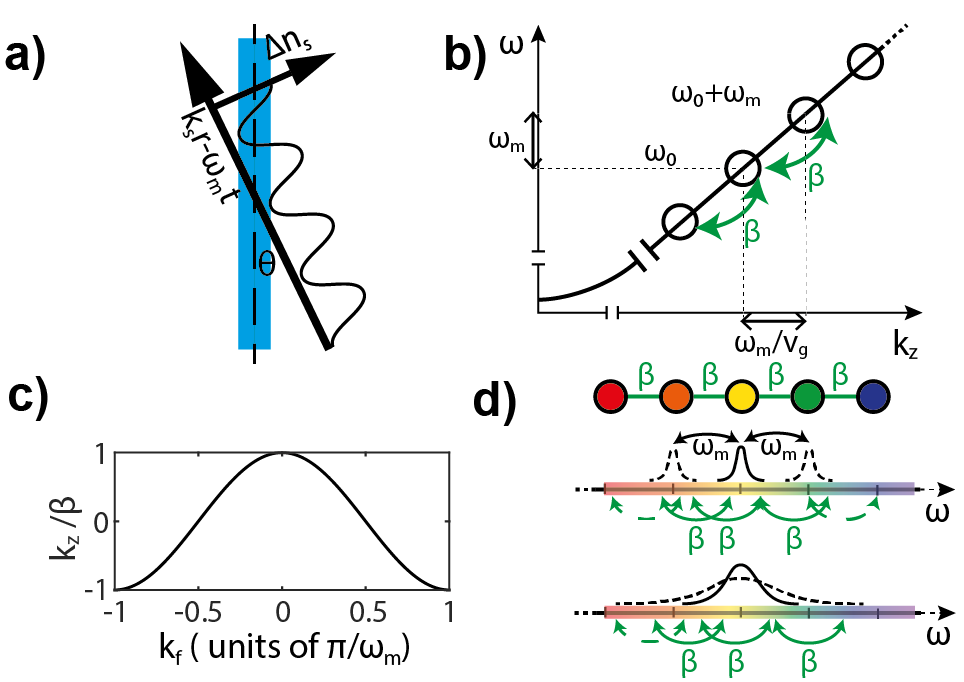}
    \caption{a) concept of the travelling sound wave used to modulate a single optical waveguide. b) Induced 1D spectral lattice in the dispersion relation of the fundamental waveguide mode. c) Dispersion relation in the frequency lattice, that is the longitudinal wavevector $k_z$ plotted as a function of the frequency momentum $k_f$. 
    d) top(bottom) Illustration of the qualitatively different dynamics for narrowband(broaband) light beam compared to the modulation frequency $\omega_m$
    }
    \label{fig:fig1}
\end{figure}

\section{Dynamical modulation of a single isolated waveguide}
Let us start from an isolated optical waveguide, whose axis is aligned to the $z$ direction. Restricting to its lowest order mode, the waveguide supports propagating states with a continuous frequency spectrum $\omega$ and propagation constant $\kappa_l(\omega)$.
Let us now include a monochromatic refractive index modulation of angular frequency $\omega_m$ traveling through the surrounding bulk material with a wavevector $\bm{k}_m$ forming an angle $\theta$ with respect to the waveguide axis (see fig.~\ref{fig:fig1}a),
\begin{equation}\label{eq:mod}
\Delta n(\bm{r},t) = 2\,\delta n\, \cos\qty(\bm{k}_m\cdot\bm{r} -\omega_m t )
\end{equation}
The effect of this monochromatic spatio-temporal modulation is to linearly couple frequency components that are separated by the modulation frequency $\omega_m$ (fig.\ref{fig:fig1}b). The equation describing the evolution of each frequency component of the light beam then takes the form of a Schr\"odinger-like equation on a continuous 1D frequency space displaying long range hoppings. Here, the light propagation direction $z$ plays the role of time and the frequency $\omega$ plays the role of a space-like variable,
\begin{equation}\label{eq:1Dcontinuous}
    i\partial_z E(z,\omega) = \frac{\Delta k}{\omega_m}\omega \,E(z,\omega) - \beta \,E(z,\omega \pm \omega_m)\,.
\end{equation}
In this equation, $\beta = \delta n\, k_0$ quantifies the hopping amplitude and the phase mismatch $\Delta k = k_m\cos\theta - \omega_m \pdv{\kappa_l}{\omega}$ between the modulation and the optical wave provides a diagonal term $\frac{\Delta k}{\omega_m}\omega E(z,\omega)$ representing a (static) scalar potential in the direction of increasing light frequency. The details of the derivation can be found in the Appendix A.

In the limiting case where the input light is either monochromatic or consists of a comb of monochromatic lines spaced by $\omega_m$, the rigorous form \eqref{eq:1Dcontinuous} of the propagation equation can be rewritten in a more familiar form. Such a condition means that the input beam has the same periodicity as the acoustic modulation of period $T_m=2\pi/\omega_m$.
In this case, the symmetry of the system allows to restrict our attention to a discrete set of frequencies $\{\omega_n\} = \omega_0 + n\omega_m$. Then, upon the definition of the frequency components $a_n(z)= E(z,\omega_n)$, the evolution \eqref{eq:1Dcontinuous} can be rewritten~\cite{Ozawa2016,Yuan2021,wu2021quasi} in terms of a discrete 1D lattice corresponding to the synthetic dimension associated to the light frequency (fig.\ref{fig:fig1}c),
\begin{equation}\label{eq:1Ddiscrete}
    i\partial_z a_n(z) = n\,\Delta k\, a_n - \beta \, a_{n\pm 1}.
\end{equation}

\begin{figure}
    \centering
    \includegraphics[scale = 1]{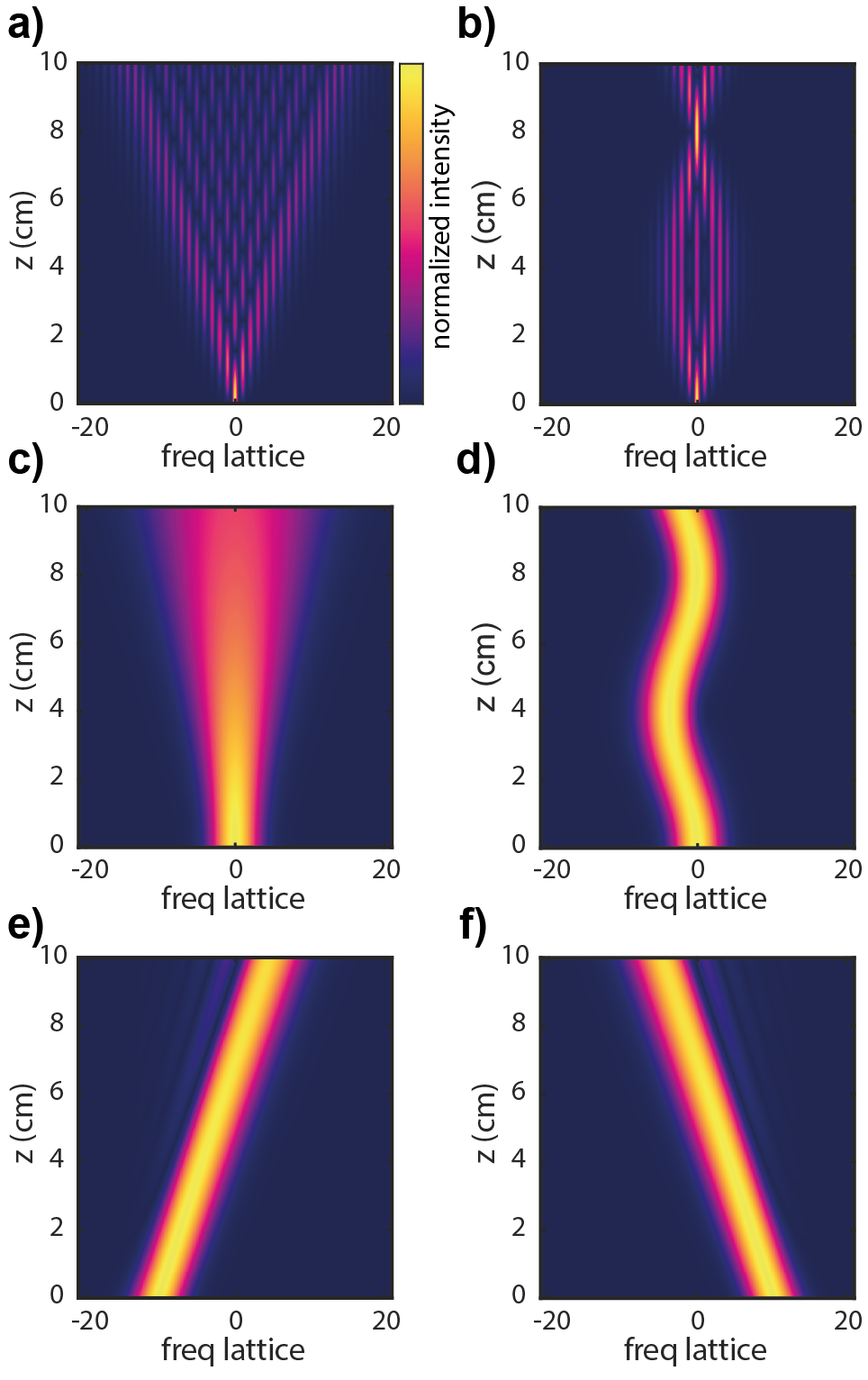}
    \caption{Power spectral density of light travelling in a dynamically modulated waveguide as a function of the propagation distance. \textbf{a,c)} Diffraction in the frequency space of a narrowband (\textbf{a}) or broadband(\textbf{c}) signal with respectively $\Delta\omega=0.3\omega_m$ and $\Delta\omega=3\omega_m$. \textbf{b,d)} Bloch oscillations in the frequency space of the same narrowband (\textbf{b}) and broadband(\textbf{d}) signals in the presence of a finite phase-mismatch $\Delta k = \beta/2$. \textbf{e,f)} Propagation in the frequency space of the broadband signal injected in the waveguide, now with a time delay $\Delta T = \pm\pi/\qty(2\omega_m)$. For this and all the following simulations $\SI{800}{\nano\metre}$ light and a $\delta n = 5\times 10^{-5}$ refractive index modulation are considered
     }
    \label{fig:fig2}
\end{figure}

While these results were known from previous works, our formalism allows our study to be straightforwardly extended to more general cases of broadband beams, for instance single optical pulses. Here, the peculiar connectivity induced by the long range hopping that is pictorially depicted in fig.\ref{fig:fig1}d makes the synthetic dimension to have an effectively discrete \eqref{eq:1Ddiscrete} or continuous \eqref{eq:1Dcontinuous} character depending on the bandwidth of the input beam. With no loss of generality, we restrict for simplicity to beams with a Gaussian spectral shape,
\begin{equation}\label{eq:narrowtime}
E_0(\omega) =e^{-\qty(\frac{\omega-\omega_0}{\Delta\omega})^2}e^{-i k_f (\omega-\omega_0)}\,.
\end{equation}
Here, the $k_f$ parameter, hereafter referred to as \textit{frequency momentum} represent the phase relation between the frequency components of the input beam and plays the same role of the momentum of a quantum particle or the transverse wavevector in real space waveguide arrays~\cite{Szameit2010}.
The eigenvalues of \eqref{eq:1Ddiscrete} are the longitudinal wavevectors of light, which play the same role of the energy of a quantum particle in a 1D lattice. In figure \ref{fig:fig1}c we plot the resulting dispersion relation, namely $k_z$, normalized to the coupling constant $\beta$, as a function of the frequency momentum.

If the spectrum of a narrowband input signal is limited within the hopping range $\Delta\omega\ll\omega_m$ (namely the light pulse duration $\tau = 2\sqrt{2\log2}/\Delta \omega$ is longer than the acoustic modulation period $T_m$), as in the top line of fig.\ref{fig:fig1}d, then all the components of such input signal excite a single cell in the synthetic space and the dynamics of the frequency spectrum recovers that of a monochromatic beam discussed in \eqref{eq:1Ddiscrete}~\cite{Qin2018_PRL,Qin2018_PRA}. 
Examples of possible behaviours are shown in fig.\ref{fig:fig2}a,b where a narrowband signal with $\Delta\omega = 0.3\omega_m$ is used as an input: In fig.\ref{fig:fig2}a, the phase mismatch $\Delta k$ is set to zero and, as a result, the spectrum envelope gets broadened by the diffraction in the discrete frequency space. On the contrary, in fig.\ref{fig:fig2}b periodic oscillations appear as a result of the linear potential induced by a finite phase mismatch $\Delta k$ and can be physically interpreted in terms of Bloch oscillations~\cite{Qin2018_PRA,Qin2018_PRL}.

A richer physics is found when the bandwidth of the injected light pulse is increased to $\Delta\omega \gg \omega_m$ (i.e. the temporal duration of the pulse decreased to $\tau \ll T_m$) and encompasses several cells. In this case, the dynamical modulation induces a coupling between different frequencies of the same broadband input signal (fig.\ref{fig:fig1}d) and the dynamics resembles the one of continuous space (fig.\ref{fig:fig2}c-f).
Most importantly, as the input signal excites multiple sites of the spectral lattice, the frequency momentum $k_f$ becomes relevant. 

Interestingly, whereas $k$-space momentum selection in CWAs is typically achieved by tilting the incident beam with respect to the waveguide axis, here the reciprocal space of the frequency dimension is the real time and $k_f$ can be understood as the time-delay ($\Delta T$) of the light pulse with respect to the sonic modulation. As a consequence, specific frequency momenta, corresponding to different group velocities in the reciprocal space, can be simply excited with an appropriate choice of the arrival time of the light pulse $\Delta T$. 

Examples of the evolution of broadband pulses with $\Delta \omega = 3\omega_m$ are displayed in fig.\ref{fig:fig2}c-f and can be physically understood in terms of the propagation constant $\varepsilon$ vs. frequency momentum $k_f$ dispersion relation plotted in fig.\ref{fig:fig1}c. In particular, in fig.\ref{fig:fig2}c, $\Delta T=0$ is chosen, which corresponds to the minimum of the dispersion relation. This results in a ballistic broadening of the  spectrum during propagation with no significant drift. In fig.\ref{fig:fig2}(d), a finite phase mismatch $\Delta k$ is instead imposed, which is responsible for Bloch oscillations in the frequency space, where the spectrum oscillates back and forth with a period along $z$ inversely proportional to the phase mismatch $\Delta k$.

For general values of $k_f$, a blue- or red-wards uniform drift is observed depending on the value of $k_f$, namely its temporal delay. Its direction and magnitude is set by the frequency group velocity $d\varepsilon/dk_f$, namely the slope of the dispersion law plotted in Fig.\ref{fig:fig1}c. The highest group velocities are obtained for frequency momenta $k_f = \pm \pi/2$ and are directed in the positive and negative directions, respectively: the corresponding $z$-dependent blue- and red-wards uniform drifts of the input signal frequency are reported in fig.\ref{fig:fig2}e,f, where we show the frequency dynamics of a light pulse injected with time delays $\Delta T = \pm\pi/2$ in a  phase-matched ($\Delta k=0$) system. Quite interestingly, the efficiency of this time-to-frequency multiplexing conversion requires the input beam to have a narrow enough distribution of frequency momenta $k_f$, that is a wide enough spectrum $\Delta \omega > \omega_m$ encompassing several cells.

An alternative point of view on this behaviour of spectrally broad and temporally short pulses is obtained by reformulating \eqref{eq:1Dcontinuous} in real-time,
\begin{equation}\label{eq:1Dcontinuous_t}
    i\partial_z e(z,t) = i\frac{\Delta k}{\omega_m}\,\partial_t e(z,t) - 2 \beta \,\cos(\omega_m t)\,e(z,t)
\end{equation}
where, as before, the role of the space-time variables is exchanged: the propagation distance $z$ plays the role of time and the physical time $t$ is a space-like variable~\cite{Larre2015}. The electric field amplitude $e(z,t)$ is here measured in the co-moving frame at the displacement speed  $v_m=\omega_m/(k_m\cos\theta)$ of the modulation profile (see Appendix A for the detailed derivation) and the temporally short light pulse occupies a small region within the modulation period.
The first term on the right-hand side, proportional to the phase-mismatch $\Delta k$, gives a uniform drift of the wavepacket along $t$. This drift can be understood as due to the wavepacket propagating at speed $v_g=\partial\omega/\partial\kappa_l$ in the laboratory frame, while the modulation wavefronts move at $v_m$.
The second term on the right-hand side can be instead understood in the analogy with a Schr\"odinger equation as a $t$-dependent potential $V_\textrm{eff}(t) = 2\beta\cos(\omega_m t)$. The derivative of the potential then provides a drift of the momentum conjugate to $t$, that is the frequency $\omega$. On the other hand, since the second order derivative term in $t$ is missing (we have set the group velocity dispersion of the waveguides to zero, $d^2\kappa_l/d\omega^2=0$), there is no kinetic energy nor any expansion of the wavepacket in $t$. 

\begin{figure*}
    \centering
    \includegraphics[scale = 1]{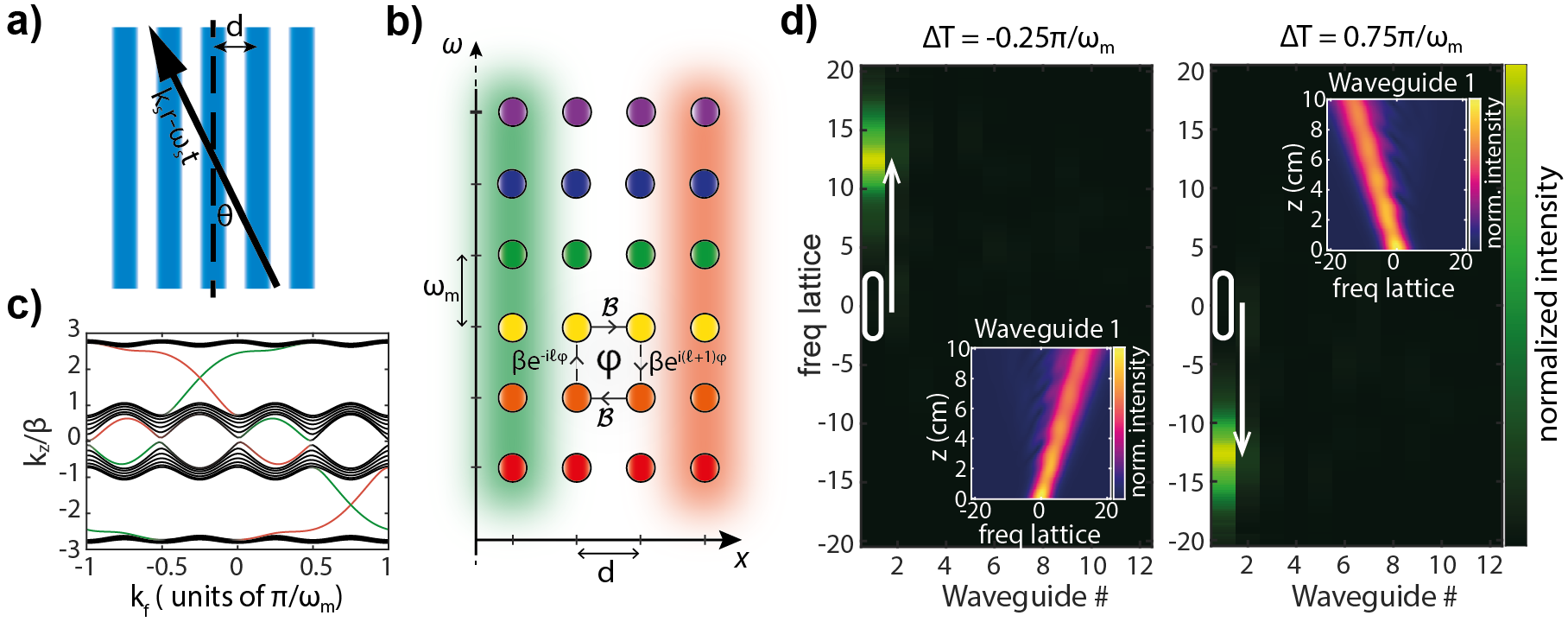}
    \caption{\textbf{a)} Scheme of a dynamically modulated array of coupled waveguides. \textbf{b)} scheme of the two-dimensional Harper-Hofstadter lattice in the hybrid space composed by the spatial and the frequency dimensions with the corresponding hopping amplitudes. \textbf{c)} Dispersion relation of a $\Phi = 1/4$ HH lattice. Edge states are color-coded accordingly to their localization as in the \textbf{b)} panel: green is for left-edge localized states and orange for right-edge. \textbf{d)} Topological frequency conversion on the right edge of the modulated waveguide array when up/down-shifting topological modes are selectively excited using properly delayed pulses. Main panels show the intensity distribution in the hybrid frequency-space of output light. The insets display the power spectral density during propagation on the excited waveguide. Same parameters as in Fig.\ref{fig:fig2}.
    }
    \label{fig:fig3}
\end{figure*}
\section{Dynamically Modulated Waveguide Arrays}
While the previous results already give a hint of the potential of modulated waveguides as a tool to control the spectrum of broadband light pulses, we are now going to see how the modulation of CWAs can capitalize on topological photonics effects to increase the robustness of the spectral manipulation.

To this purpose, let us now include a physical dimension in the model: consider a 1D array of identical waveguides distributed along the $x$-direction separated by a constant distance $d$ 
(fig.\ref{fig:fig3}a). For a sonic modulation propagating at an angle $\theta$ with respect to the waveguide axis, the dynamical modulation can be written in terms of the $\ell$-th waveguide position $x_\ell = \ell d$ as
   $\Delta n_\ell(z, t) = 2b\cos(k_m^z z + k_m^x \ell d -\omega_m t)\,,$
where $k_m^z = k_m\cos(\theta)$ and $k_m^x = k_s\sin(\theta)$ are, respectively, the longitudinal and orthogonal components of the modulation wavevector with respect to the waveguides axis. 

Analogous calculations with respect to the single waveguide case (see the Appendix A for details of the derivation) lead to the following set of propagation equations for the modal amplitude in the $\ell$-th waveguide at frequency $\omega$
\begin{multline}\label{eq:2d}
i\partial_z E_\ell(z,\omega) = \frac{\Delta k}{\omega_m}\,\omega \,E_\ell(z,\omega) \\ - \beta e^{\pm i \ell \Phi}\,E_\ell(z,\omega\mp\omega_m)-\mathcal{B}\,E_{\ell\pm 1}(z,\omega)
\end{multline}
where $\beta$ is again the coupling coefficient in the frequency dimension, $\mathcal{B}$ is the coupling coefficient in the spatial dimension which encapsulates the modal overlap between neighboring waveguides, $\Delta k$ is the usual phase mismatch term and $\Phi = k_m d\sin(\theta)$.

For a narrowband incident pulse, it is again possible to discretize the frequency dimension by restricting to $\{\omega_n\} = \omega_0 + n\omega_m$. This leads to a 2D rectangular lattice (fig.\ref{fig:fig3}b) in a hybrid space composed by a spatial and a frequency dimension and to a propagation equation:
\begin{equation}\label{eq:2ddiscrete}
\begin{split}
i\partial_z a_{\ell,n} = n\,\Delta \,k \,a_{\ell,n} - \beta \,e^{\pm i \ell \Phi}\,a_{n\mp 1,\ell}-\mathcal{B}\,a_{n,\ell\pm 1}
\end{split}
\end{equation}
which takes the form of a Harper-Hofstadter (HH) model~\cite{Ozawa2019}. Because of the non-reciprocal phase term in the frequency-space coupling, proportional to $\beta$, a loop around a single plaquette of the rectangular lattice is in fact associated to a phase shift of $\Phi = \oint_c A\cdot\dd \bm r = k_md\sin(\theta)$. This phase shift is analogous to the geometric phase acquired by a charged  particle in a constant magnetic field 
piercing the lattice.

\subsection{Topological 2D physics in the synthetic space}
For a rational $\Phi={p}/{q}$ (with $p,q$ co-prime integers), the band-structure of the HH model is composed by $q$ bands with non-zero Chern numbers signalling non-trivial topological properties~\cite{Ozawa2019}. As a consequence of such non trivial topology, in a finite lattice geometry, chiral edge states appear within each bandgap and can be used to unidirectionally transport light around the edge of the system. 
In particular, we consider a ribbon geometry with a finite extension along the spatial dimension and infinite in the frequency dimension. The resulting bandstructure is shown in fig.\ref{fig:fig3}c: depending on the value of the frequency momentum $k_f$, mid-gap states appear with well defined and opposite group velocities, exponentially localized at the opposite spatial edges of the waveguide array, indicated by the green and orange curves in the figure. Since a single state only exists in each gap for each edge, they are immune to elastic back-scattering on disorder and allow for unidirectional transport, in our case along the frequency direction.

To demonstrate that unidirectional states with a specific chirality can be excited in a realistic experiment we inject a broadband input state with specific values of the frequency momenta $k_f$ into the left-most waveguide of the array. As already described in the single waveguide case, the frequency momentum $k_f$ corresponds to the time delay $\Delta T$ with respect to the dynamical modulation: having a localized beam in $k_f$ then corresponds to having a temporally short pulse, well shorter than the modulation period $T_m$ and wider in frequency than the modulation frequency $\omega_m$.

Fig.\ref{fig:fig3}d shows how this procedure indeed results in the excitation of blue- or red-shifting topological modes, depending on the excited edge and on the value of the frequency momentum: as one can see from the dispersion relation of the specific HH lattice under consideration, each edge supports indeed a single edge state for each value of $k_f$: thanks to this fact, there is no need to use additional ``straw'' waveguides to select states with the desired propagation constant~\cite{stutzer2018,wu2021quasi}. As this topological state is exponentially localized at the interface, light stays confined in the vicinity of the spatial edge and does not substantially penetrate into the bulk waveguide, whereas the overall frequency spectrum is shifted towards higher (or lower) frequencies, as illustrated in the inset.

\begin{figure}[htbp]
    \centering
    \includegraphics[scale = 1]{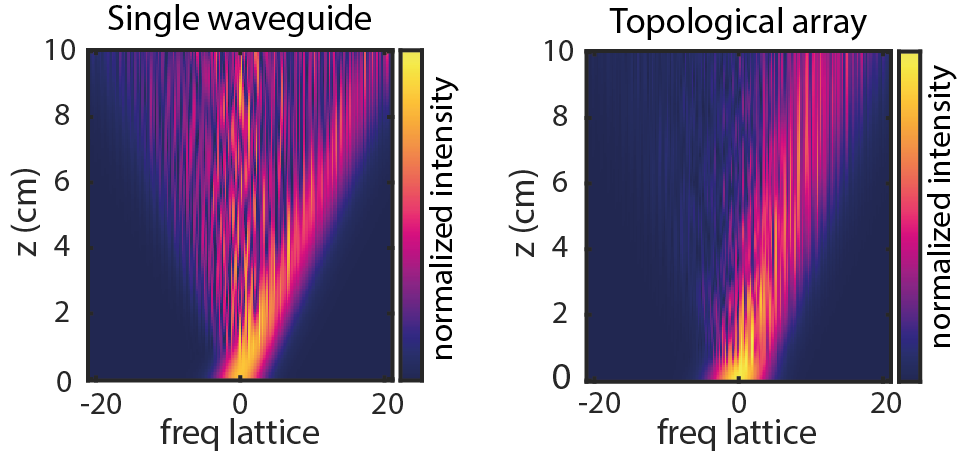}
    \caption{ Spectral power density distribution as a function of the propagation distance in the presence of a frequency-dependent refractive index disorder for a broadband beam propagating in a single modulated waveguide (left) or in a unidirectionally propagating  edge mode of a topological array (right).
    Same parameters of Fig.\ref{fig:fig2}, the noise is uniformly distributed with maximum amplitude $\Delta n\approx \beta/k_0$
    }
    \label{fig:fig4}
\end{figure}

The key novelty of the topologically protected frequency transport in the hybrid frequency-space lattice is highlighted in fig.\ref{fig:fig4}. While in the clean system the frequency shift may look very similar to the one of the isolated waveguide configuration shown in figure \ref{fig:fig2}(e,f), the behaviour in the presence of disorder is completely different. In the figure, we consider as an example a frequency-dependent refractive index, as it would be the case with an impure optical medium.
For an isolated waveguide, such a disorder would induce a localization effect, so that the frequency conversion is inhibited (fig.\ref{fig:fig4}a). On the contrary, as shown in fig.\ref{fig:fig4}b, when happening through a topological mode the frequency shift is robust. 

This result demonstrates the power of hybrid frequency-space lattices for shifting the carrier frequency of a broadband signal in a topologically protected way. In particular, the direction and the magnitude of the frequency shift
can be controlled by the temporal delay $\Delta T$ of the incident pulse and the wavevector of the dynamical modulation. From the point of view of topological photonics, an exciting asset of our proposal consists in the possibility of varying the value of the synthetic magnetic field $\Phi$ through the properties of the sound wave and, in this way, span across models with different topological properties. Further manipulation tools are offered by the dispersion relation of the different waveguides~\cite{Oliver2021} and/or the frequency-dependence of their refractive index~\cite{Ozawa2016}, which respectively induce external potentials along the spatial and frequency directions.

\subsection{Edge Effects in Frequency space}
While a fundamental limitation to the size of the lattice is imposed by the finite physical size of the waveguide array, its size along the the frequency dimension is effectively unbounded. A higher order expansion of the waveguide dispersion relation, taking into account the group velocity dispersion $\partial^2{\kappa_l}/{\partial\omega^2}$ of the waveguides~\cite{Larre2015} would induce a parabolic confining potential in the frequency dimension, but this effect is tipically only relevant when sizable frequency shifts are considered, which is not the case of this first study. A detailed discussion of these higher order effect will be the subject of a future work.

In this section we follow an alternative route and we show how absorbing elements with a strongly localized absorption spectrum (due for instance to impurities in the constitutive material of the optical waveguides) can act as a hard edge in the frequency dimension, as first proposed in~\cite{Ozawa2016}, with interesting applications to filtering or frequency routing. 

We consider a frequency-dependent absorption with a Gaussian profile, centered at a frequency $\omega_a$ with bandwidth of $\Delta\omega_a$. Such an absorption is modelled as an additional frequency dependent term $\alpha(\omega)$ in the equation of motion \eqref{eq:2d}:
\begin{equation}
\begin{split}
i\partial_z E_\ell(z,\omega) = \qty(\frac{\Delta k}{\omega_m}\,\omega + \alpha\qty(\omega))E_\ell(z,\omega) \\ - \beta e^{\pm i \ell \Phi}\,E_\ell(z,\omega\mp\omega_m)-\mathcal{B}\,E_{\ell\pm 1}(z,\omega),
\end{split}
\end{equation}
where we take for concreteness
\begin{equation}
\alpha\qty(\omega) = -i\,\bar{\alpha}\, e^{-\qty(\frac{\omega-\omega_a}{\Delta \omega_a})^2}
\end{equation}
with a strong $\bar{\alpha} \gg \beta$. This latter condition enforces a Zeno regime where absorption is effectively blocked and negligible power is dissipated\cite{Peres1980}.
Because of the peculiar long range connectivity that characterizes the synthetic frequency dimension, the behavior of the system qualitatively changes according to the bandwidth of the absorbing impurity. 

\begin{figure}[h]
    \centering
    \includegraphics[scale = 1]{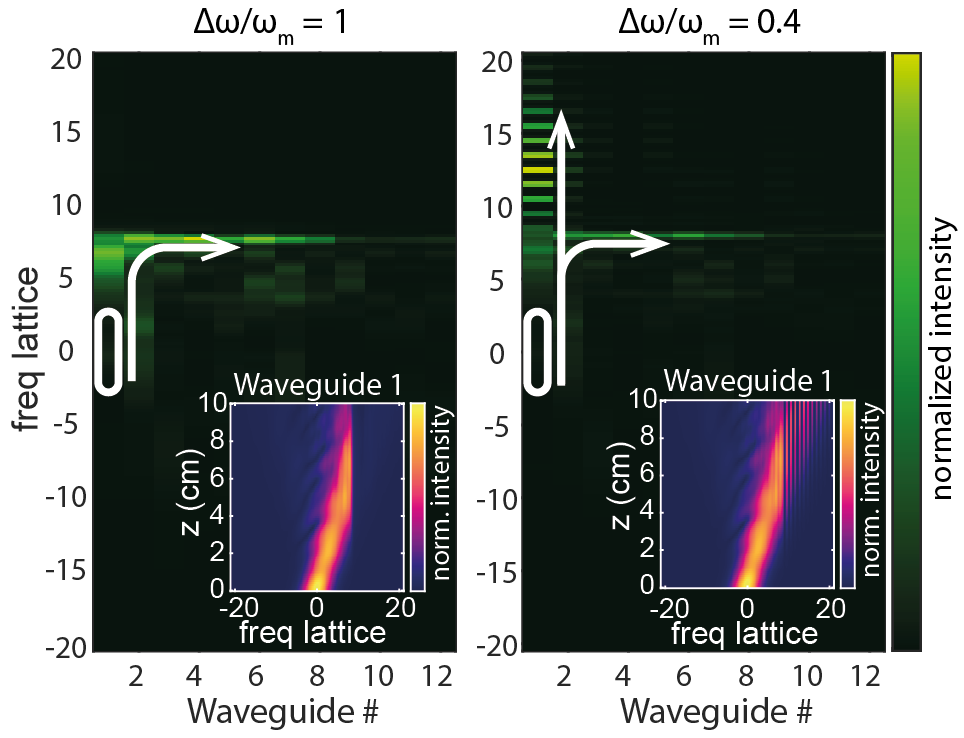}
    \caption{Spectral power density distribution in each waveguide after a propagation of \SI{10}{\centi\metre} when a strong absorption, centered at ${\omega}/{\omega_m} = 9$, is considered. \textbf{Left:} Broadband absorption acts as an effective hard-wall for the whole wavepacket. \textbf{Right:} Narrowband absorption acts as an effective hard wall only for a fraction of the wavepacket spectrum, while the rest keeps propagating undisturbed. \textbf{Insets:} spectral power density distribution as a function of the propagation distance in the leftmost waveguide of the array.
    }
    \label{fig:fig5}
\end{figure}

Let's first consider a broad absorption spectrum $\Delta\omega_a\gg\omega_m$, such that all the frequency components of an incident wavepacket will likewise experience the hard edge in frequency space. As a result, as no counterpropagating states at the same propagation constant $\varepsilon$ exist, the entire wavepacket will be deflected with perfect transmission around the corner of the hybrid space and will keep propagating along the physical direction. This behavior is illustrated in fig.~\ref{fig:fig5}~left, where a broadband light pulse is injected in the leftmost waveguide of a dynamically modulated array with the right frequency-momentum $k_f$ to excite a blue-shifting topological mode. As soon as the wavepacket hits the broadband absorption the blue-shift stops and light is unidirectionally guided along the spatial dimension. 

On the contrary, if the absorption spectrum is narrowband ($\Delta \omega_a\ll\omega_m$) the long range interaction will allow for some frequency component to freely go across the absorption band. This behavior is illustrated in fig.~\ref{fig:fig5}~right. There, a blue-shifting wavepacket hits a narrowband absorption while propagating along the frequency dimension. When they collide with the frequency wall, a series of bins corresponding to the absorption band are chopped out from the wavepacket spectrum and are deflected along the physical dimension. The rest of the wavepacket does not interact with the absorption band and keeps propagating along the frequency dimension in the form of the modulated up-shifting frequency spectrum visible in the inset. 

This unique feature is the result of topological physics (unidirectional reflection-less propagation) for broadband beams in a continuous frequency synthetic dimension and does not have, to the best of our knowledge, any counterpart in cavity-based topological schemes. With a suitable design of the central frequency and bandwidth of the absorption band, such effect could find interesting applications in telecommunication technology such as frequency filtering, switching and (de)multiplexing. 

\section{Conclusions and perspectives} In this work we have proposed a scheme to realize a hybrid frequency-space synthetic lattice subject to a synthetic magnetic field by acousto-optically modulating an array of waveguides. The strength of the synthetic magnetic field is controlled by the wavevector of the sound wave and, in turn, determines the topological properties of the lattice. Specific values of the frequency momentum in the synthetic direction can be selected by tuning the arrival time of the incident pulse with respect to the modulation period, leading to an efficient time-to-frequency multiplexing conversion.

The proposed scheme paves the way to the use of the standard photonic technology of laser-written waveguides for the spectral manipulation of broadband signals. The coherent nature of the proposed scheme permits its application both to classical waves and to single- or few-photon quantum wavepackets, e.g. to manipulate their entanglement between time, frequency and spatial variables. This opens the way to various applications in telecommunications and quantum science and technology.

\acknowledgments{}
FSP acknowledges fundings from Deutscher Akademischer Austauschdienst (grant n. 57507869). IC and FSP acknowledge financial support from the European Union H2020-FETFLAG-2018-2020 project ``PhoQuS'' (n.820392) and from the Provincia Autonoma di Trento, partly through the Q@TN initiative. AS acknowledges funding from the Deutsche Forschungsgemeinschaft (grants SCHE 612/6-1, SZ 276/12-1, BL 574/13-1, SZ 276/15-1, and SZ 276/20-1) and the Alfried Krupp von Bohlen and Halbach Foundation. FSP thanks Alberto Nardin for fruitful discussions.

\appendix
\section{Derivation of the equations of motion}
In the following we provide further details about the derivation of the equation of motion in the hybrid space-frequency lattice

\subsection{Equation of motion for a single modulated waveguide}
Let's consider paraxial light propagation in an optical waveguide with the axis aligned to the $z$-direction. Under the assumption of slowly varying refractive index modulation, the modal amplitude of the electric field is also assumed to vary slowly, and the paraxial Helmholtz equation can be rewritten as a Schr\"dinger-like equation
\begin{equation}\label{eq:singlewaveguide}
    i\partial_z \mathcal{E}(z,t) = -k\, \mathcal{E}(z,t) -\Delta n\, k_0\, \mathcal{E}(z,t)
\end{equation}
Here $k$ is the light wavevector that is a function of the temporal derivative and the dispersion. We consider a refractive index perturbation that has a travelling monochromatic wave profile such as
\begin{equation}
    \Delta n(r,t) = 2 \,\delta n \,\cos\qty(\bm {k_m} \cdot \bm r - \omega_m t)
\end{equation}
Since the typical optical waveguide width is much smaller than the wavelength of the perturbation, we assume the perturbation phase to be constant in the transverse plane therefore we can rewrite the modulation as
\begin{equation}
    \Delta n(z,t) =2\,\delta n \,\cos\qty(k_m^z z -\omega_m t)
\end{equation}
where $k_m^z = k_m\cos(\theta)$, being $\theta$ the angle of the modulation wave with respect to the waveguide axis, is the z-component of the modulation wavevector. We move to the frequency space and rewrite \eqref{eq:singlewaveguide} as
\begin{equation}
    i\partial_z  \mathcal{E}(z,\omega) = -\kappa_l(\omega) \, \mathcal{E}(z,\omega) - k_0 \,\Delta n (z,\omega) \circledast  \mathcal{E}(z,\omega)
\end{equation}
Where $\circledast$ denotes convolution in the frequency space, $\Delta n(z,\omega) = \delta n e^{\pm i k_m^z z}\delta(\omega \mp \omega_m)$ is the Fourier Transform of $\Delta n(z,t)$ and $k(\omega)$ is the light wavevector that we can expand around $\omega_0$ as $k(\omega) = k_0 + \frac{\omega-\omega_0}{v_g}$, being $v_g$ the group velocity in the waveguiding struture.
We now perform a gauge transformation and move to the light wave reference frame by defining
\begin{equation}\label{eq:gaugetransform1}
     \mathcal{E}(z,\omega) = \tilde{\mathcal{E}}(z,\omega)e^{ik(\omega)z}
\end{equation}
with this substitution the equation for $\tilde{\mathcal E}(z,\omega)$ reads
\begin{equation}
    i\partial_z \tilde{\mathcal{E}}(z,\omega) = -k_0\,\delta n\, e^{\pm i\qty(k_m - \frac{\omega_m}{v_g})z}\tilde{\mathcal{E}}(z,\omega \mp\omega_m)
\end{equation}
In which we used the fact that $k(\omega\pm\omega_m) - k(\omega) = \pm{\omega_m}/{v_g}$. We define the phase mismatch between the light wave and the perturbation as $\Delta k = k_m -{\omega_m}/{v_g}$ and perform another gauge transformation
\begin{equation}\label{eq:gaugetransform2}
    \tilde{\mathcal{E}}(z,\omega) = E e^{i\frac{\omega}{\omega_m}\Delta k z}
\end{equation}
which allows to write the equation for E:
\begin{equation}\label{eq:1Dcontinuous_SM}
    i\partial_z E = \frac{\Delta k}{\omega_m}\,\omega \,E - \beta\, E(\omega\mp\omega_m)
\end{equation}
Where $\beta = k_0 \delta n$ is a coupling coefficient. The diagonal term $\frac{\Delta k}{\omega_m}\omega E(z,\omega) = V \omega E(z,\omega)$ plays the role of an on-site energy increasing in the direction of increasing frequencies, which is the analogue of a constant scalar potential.\\
The scalar potential $V$ is proportional to the phase mismatch: 
\begin{equation}\label{eq:phasemismatch}
   V \propto \Delta k = k_m^z-\omega_m\pdv{k}{\omega} =k_m\cos\theta - \frac{\omega_m}{v_g}
\end{equation}
where $v_g \approx c_0/n$ is the group velocity of light in the waveguide. Here,  the phase matching condition corresponds to having $\Delta k = 0$ in \eqref{eq:phasemismatch}. 

As the light wavevector is extremely small, one can typically assume $\Delta k \approx k_m\cos\theta$. However $k_m = {2\pi\omega_m}/{c_m}$, where $c_m$ is the speed of the modulation wave is a significant quantity. As a result the slightest deviation from the optimal propagation angle $\theta = \pm\pi/2$ makes the phase mismatch too strong to observe any dynamics in the synthetic space at all. As we further discuss in the implementation section this motivates the need of precisely controlling the direction of the light wave which is a notable experimental effort.

\subsection{Equation of motion for a modulated waveguide array} 
Let's now consider an array of modulated waveguides. If we consider no coupling between the individual waveguides then each waveguide of the array can be described by an equation analogous to \eqref{eq:singlewaveguide}. On the other side, if the waveguides are weakly coupled \eqref{eq:singlewaveguide} is modified to include a coupling coefficient between the waveguides yielding the following set of equations for the $\ell$-th waveguide of an array
\begin{equation}\label{eq:waveguidearray}
\begin{split}
i\partial_z\mathcal{E}_\ell(z,t)_\ell = -k \,\mathcal{E}_\ell - k_0\,\Delta n (z,t)\,\mathcal{E}_\ell(z,t)\\
-\mathcal{B} \,\mathcal{E}_{\ell\pm 1}(z,t)
\end{split}
\end{equation}
Where we assumed the waveguides to be all identical such that $k_\ell = k$ and the hopping amplitude is $\mathcal{B}$.\\
The refractive index perturbation, as previously considered, is a travelling wave described by
\begin{equation}
    \Delta n(z,t) = 2\,\delta n\, \cos(\bm{k_m}\cdot\bm{r} - \omega_m t)
\end{equation}
We consider the waveguides to be evenly distributed along the $x$-direction with an inter-waveguide distance $d$ that we assume to be much larger than the waveguide width. The phase of the perturbation is therefore different in each individual waveguide, but can be considered constant across the width of each waveguide. To this aim we discretize the space in the $x$-direction and define the position of the $\ell$-th waveguide as $x_\ell = d\ell$. The refractive index perturbation can be rewritten as
\begin{equation}
    \Delta n_\ell(z,t) = 2\,\delta n\, \cos(k_m^z z + k_m^x d\ell - \omega_m t)
\end{equation}
where $k_m^z = k_m\cos(\theta)$ is the longitudinal component while $k_m^x = k_m\sin(\theta)$ is the transverse component of the modulation wavevector. Following the same steps of the single waveguide case we move to the frequency space and rewrite \eqref{eq:waveguidearray} as
\begin{equation}
\begin{split}
i\partial_z \mathcal{E}_\ell(z,\omega) = -k(\omega) \,\mathcal{E}_\ell(z,\omega) - k_0\,\Delta n_\ell (z,\omega)\,\mathcal{E}_\ell(z,t) \\
- \mathcal{B}\, \mathcal{E}_{\ell\pm 1}(z,\omega)   
\end{split}    
\end{equation}
In this case $\Delta n(z,\omega)$ is the Fourier transform of $\Delta n(z,t)$ and is written as $\Delta n_\ell(z,\omega) = e^{\pm i \qty(k_m^z z + k_m^x \ell d)}\delta(\omega \mp \omega_m)$. We apply to each waveguide the gauge transformation defined in \eqref{eq:gaugetransform1} and rewrite \eqref{eq:waveguidearray} as:
\begin{equation}
\begin{split}
i\partial_z \tilde{\mathcal{E}}_\ell(z,\omega) = -\beta \,e^{\pm i\Delta kz}\,e^{\pm i \Phi \ell}\,\tilde{\mathcal{E}}_\ell(z,\omega \mp\omega_m) \\
- \mathcal{B} \,\tilde{\mathcal{E}}_{\ell \pm 1}(z,\omega)
\end{split}
\end{equation}
Where $\Delta k = k_m\cos\theta - \frac{\omega_m}{v_g}$ is the usual phase mismatch term, $\Phi = dk_m\sin\theta$ is the modulation phase difference between two closest waveguides, $\beta$ is the coupling coefficient in the frequency space and $\mathcal{B}$ is the coupling coefficient in the physical space. Using once again \eqref{eq:gaugetransform2} on each waveguide we can rewrite the latter as
\begin{equation}\label{eq:2Dcontinuous}
\begin{split}
i\partial_z E_\ell(z,\omega) = \frac{\Delta k}{\omega_m}\omega E_\ell(z,\omega)-\beta \,e^{\pm i \Phi \ell}\,E_\ell(z,\omega \mp\omega_m) \\
- \mathcal{B} \,E_{\ell \pm 1}(z,\omega)
\end{split}
\end{equation}

\section{Notes on the experimental implementation}
Our model takes advantage of the continuous spectrum of modes in waveguides and allows to effectively manipulate a TLC signal with arbitrary frequency spectrum. Previous works have been focused on the modulation of isolated LiNbO{\scriptsize 3} waveguides using microwaves and the electro optical effect \cite{Qin2018_PRA, Qin2020}. However LiNbO{\scriptsize 3} waveguides require complex fabrication procedure and the technology is not suited for the realization of coupled waveguide arrays, as the strong confinement of the light field strongly suppresses the evanescent coupling between adjacent waveguides. Furthermore, refractive index variation via RF modulation requires a strong electro-optical coefficient which is available in few materials and especially not in amorphous ones, that are more suited for the realization of coupled waveguides arrays \cite{Szameit2010}
As a different and more promising paradigm we consider here the use of acoustic waves to perform a dynamical modulation of optical waveguide arrays fabricated using femtosecond laser writing in SiO{\scriptsize 2} glass \cite{Szameit2010}.

\begin{figure}[htbp]
    \centering
    \includegraphics[scale = 1]{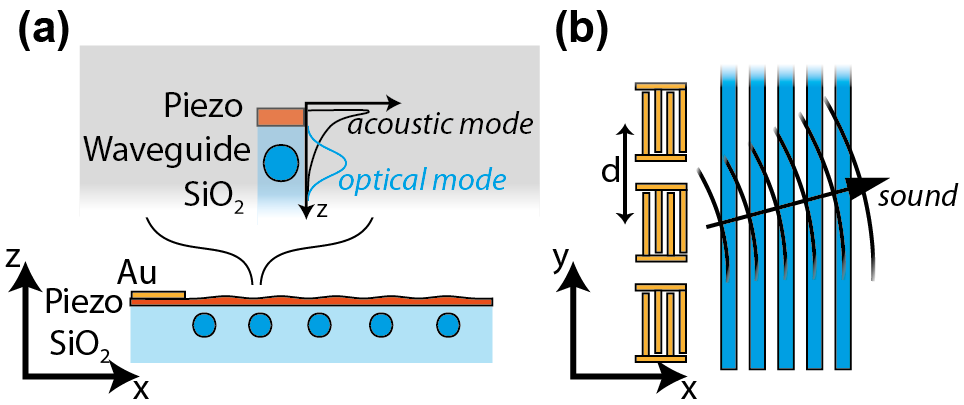}
    \caption{Schematic of the proposed experiment. \textbf{(a)}: Front view of the device. A thin layer of a piezoelectric material is deposited on a SiO{\scriptsize 2} sample with surface waveguides. Inset: The evanescent tail of the Rayleigh wave supported by the piezo layer leaks into the fused silica waveguide modulating the refractive index via the acousto-optic effect. \textbf{(b)}: Top view of the device. InterDigitalized Transducers (IDTs) with a fixed distance $d$ are lithographed on the sample surface.}
    \label{fig:SI1}
\end{figure}

The numerical calculations presented in the main text assume a modest refractive index increase of $2\cdot10^{-5}$ that, using a \SI{100}{\mega\hertz} acoustic wave generates frequency shift of $\approx \SI{800}{\mega\hertz}$ after a $\SI{10}{\centi\metre}$ propagation into a single modulated waveguide. Higher frequency shifts are achievable using higher frequency acoustic waves, up to the point where higher order phenomena becomes important such as the dispersion of the coupling coefficient in the frequency space or the parabolic confining potential in the frequency space caused by group velocity dispersion. 

However the higher the frequency of the dynamical modulation the more dramatic is the effect of phase mismatch,  thus a mechanism to enforce directional excitation of the sound wave with high precision is required. To this aim, we suggest to excite the sound wave using an array of emitters. On one hand multiple emitters allow to shape the front of the acoustic wave, thus getting closer to the ideal plane wave front. On the other hand, the use of multiple equi-spaced emitters with an externally controllable phase relation allows to effectively control the propagation angle of the modulation. The ability to electronically control the phase relation between different emitters would also allow the realization of reconfigurable devices.

Previous works have inspected the propagation of pressure waves in bulk fused silica samples \cite{Poffo2009} characterizing the acousto-optical effect on ion-diffused waveguide structures in a low acoustic-frequency regime. Despite the attractiveness of bulk acoustic waves, that would allow to perform dynamical modulation of two dimensional waveguide arrays hence potentially achieving an effective 3D physics \cite{Qin2018_OptExpr}, the power decay of spherical waves is too fast to allow for an homogeneous modulation of large waveguide arrays. 

To overcome this issue, in the recent years, a strong effort has been devoted to the integration of Surface Acoustic Waves (SAW) technologies \cite{delsing2019}. 
Being confined to the surface, the power scaling with the travelled distance is in fact much slower, allowing to conserve the acoustic power throughout the entire sample length. Moreover, SAW transducers can be realized with extremely high operating frequencies \cite{delsing2019,Tadesse2014}, leading to bigger frequency shifts in our proposed modulator.  \par 
SAW are typically excited using InterDigitalized Transducer (IDTs), whose fabrication is compatible with standard lithographic process and is particularly suited for the realization of emitter arrays~\cite{delsing2019}. In particular, in our proposed scheme, IDTs arrays could be lithographed on the fused silica glass after sputtering the surface of the sample with a piezoelectric film as in figure \ref{fig:SI1}. 
Although in this scheme the SAW is peaked in the thin piezoelectric layer, if the optical waveguides are realized sufficiently close to the interface a reasonable overlap is expected \cite{Tadesse2014} between the evanescent tail of the acoustic wave (in the order of an acoustic wavelength) and the optical mode field, leading to the required effective index modulation.
\end{document}